\documentclass[11pt,twoside]{article}
\usepackage{asp2010}

\resetcounters

\begin{document}

\title{Modelling the recurrent nova U Scorpii in quiescence}
\author{M. P. Maxwell,$^1$ M. T. Rushton,$^1$ and S. P. S. Eyres$^1$}
\affil{$^1$Jeremiah Horrocks Institute, University of Central Lancashire, Preston, UK, PR1 2HE}

\begin{abstract}
VLT and SALT spectroscopy of U Sco were obtained $\sim$18 and $\sim$30
months after the 2010 outburst. From these spectra the accretion disc
is shown to take at least 18 months to become fully reformed. The
spectral class of the companion is constrained to be
F8$^{+5}_{-6}$\,IV-V at the 95\% confidence level
when the irradiated face of the companion is visible.
\end{abstract}

\section{Introduction}
The recurrent nova U Scorpii was observed to go into outburst in
January 2010. The peak and decline to the pre-outburst V band
magnitude were very well observed \citep{schaeferdisc}, however there
are very few observations of the system at the quiescent magnitude level (see \citealt{johnston} for an example). U~Sco is a semi-detached binary with a high mass white dwarf primary \citep{thoroughgood} and an orbital period of 1.23 days \citep{schaefer95}. The companion is a sub-giant with a mass of 0.88\,M$_{\odot}$ and radius 2.1\,R$_{\odot}$ \citep{thoroughgood}. 

\section{Observations}
\begin{figure}
\includegraphics[width=\textwidth]{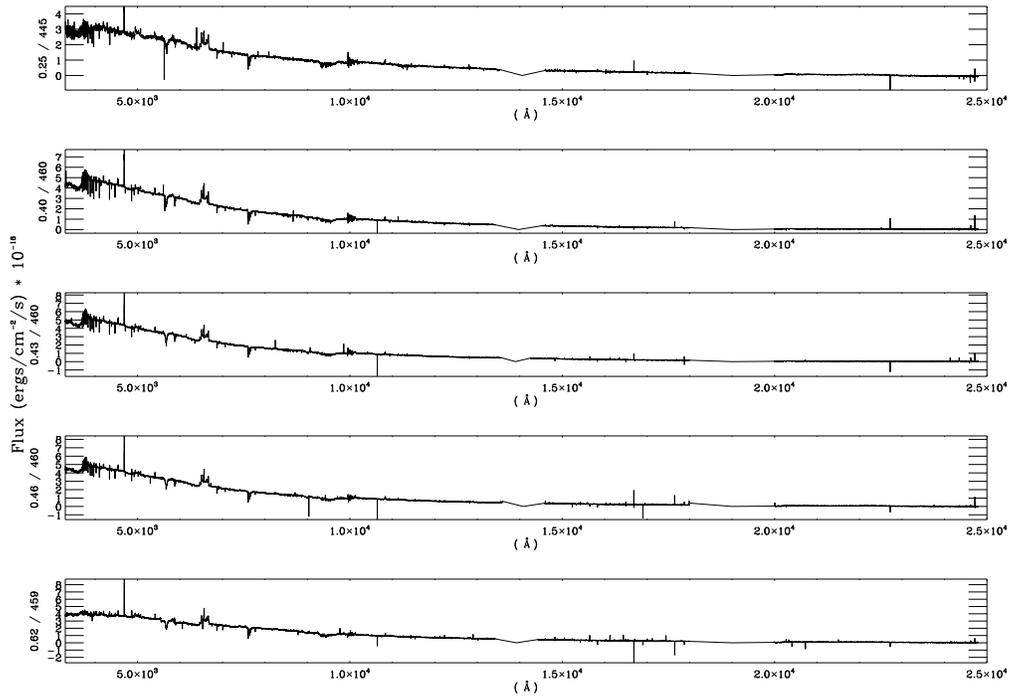}
\caption{VLT spectra. Orbital phase and number of days after outburst are shown on the y axis.}
\label{vltspecs}
\end{figure}

\begin{figure}
\includegraphics[width=\textwidth]{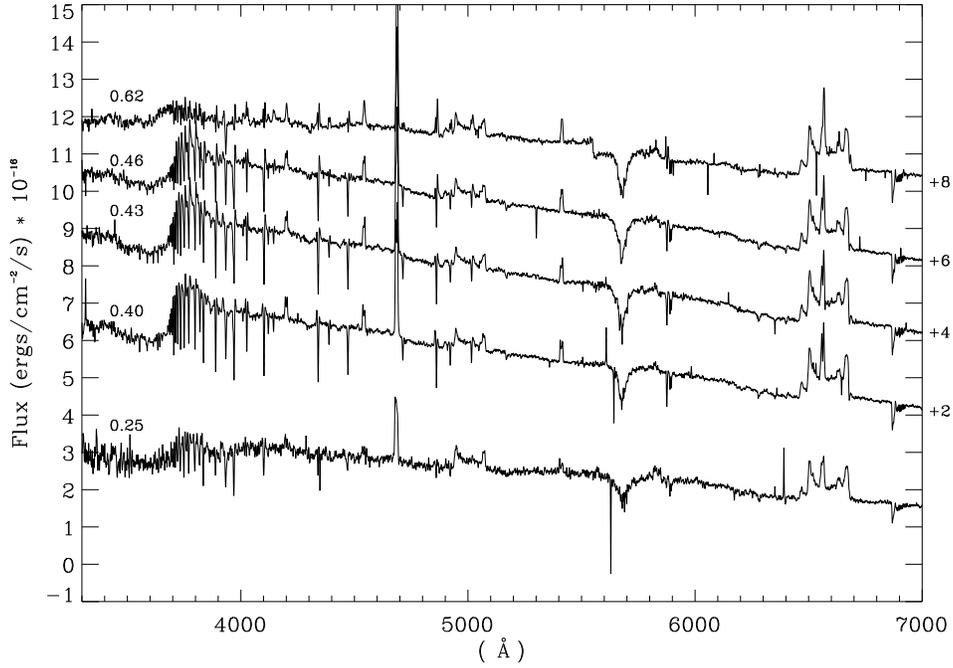}
\caption{VLT spectra from 3300 - 7000{\AA}, offset as shown to the right of the Figure.}
\label{zoomspecs}
\end{figure}

\begin{figure}
\includegraphics[width=\textwidth]{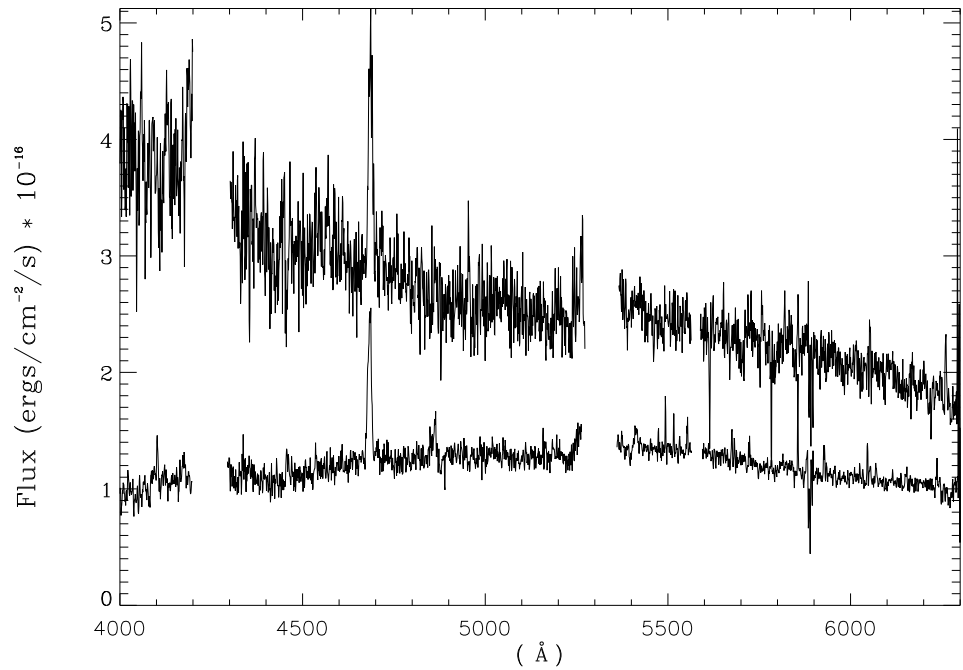}
\caption{SALT spectra taken at phase 0.55, day 848 (lower spectrum) and phase 0.77, day 909 (upper spectrum).}
\label{saltspecs}
\end{figure}

Following the 2010 outburst of U~Sco and the subsequent return to the
pre-outburst magnitude, VLT and SALT spectra were obtained $\sim$18
and $\sim$30 months after outburst respectively, as detailed in Table
\ref{obslog}. The VLT spectra, R $\sim$5000, can be seen in Figure
\ref{vltspecs}, with the optical region shown in Figure
\ref{zoomspecs}. SALT spectra, R $\sim$1000, are displayed in Figure
\ref{saltspecs}. The VLT spectra show several emission lines, the
strongest of which is He\,{\sc ii} 4686{\AA}. H Balmer lines are
present and have profiles consistent with those expected from an
optically thick accretion disc \citep{warner}. 

The VLT spectra. taken $\sim$18 months after outburst, are clearly
different from those taken 2-3 years after outburst by
\cite{johnston}, however they show many of the same features observed
by \cite{thoroughgood} $\sim$50 days after the 1999 outburst. The SALT
spectra, taken $\sim$30 months after outburst, appear much more
similar to the spectra of \cite{johnston}. Clearly there is
significant spectroscopic evolution for several months after the
system has returned to the pre-outburst magnitude.

\begin{table}
\centering
\begin{tabular}{cccc}\hline
Orbital phase & Days after outburst & Facility & Wavelength range ($\mu$m)\\ \hline
0.25 & 445& VLT& 0.3 - 2.5\\
0.40 & 460& VLT& 0.3 - 2.5\\
0.43 & 460& VLT& 0.3 - 2.5\\
0.46 & 460& VLT& 0.3 - 2.5\\
0.55 & 848& SALT& 0.4 - 0.63\\
0.62 & 459& VLT& 0.3 - 2.5\\
0.77 & 909& SALT& 0.4 - 0.63 \\ \hline

\end{tabular}
\caption{Observing Log.}
\label{obslog}
\end{table}

\section{The accretion disc}
The nature of the accretion disc in U Sco was constrained by fitting a
model to the VLT spectra. There are two components to the model, one
representing the accretion disc and the other the companion star,
which is represented by a single model stellar atmosphere. The
accretion disc model consists of several annuli, each represented by a
model stellar atmosphere produced using the ATLAS code by
\cite{kurucz}. The effective temperature of each annulus was determined using

\begin{equation}
T_{eff} = T_{*}\left(\frac{R_{acc}}{r}\right)^{3/4} \left(1 - \left(\frac{R_{acc}}{r}\right)^{1/2}\right)^{1/4},
\label{radial}
\end{equation}
where R$_{acc}$ is radius of the white dwarf, r is the distance from
the white dwarf to the disc annulus, and T$_{*}$ is found using

\begin{equation}
T_* = 4.10 \times 10^4 R_9^{-3/4} M_1^{1/4} \dot{M}_{16}^{1/4},
\label{tstar}
\end{equation}
where R$_9$ is the white dwarf radius in units of 10$^9$cm, M$_1$ is
the white dwarf mass in solar masses, and $\dot{M}_{16}$ is the mass
accretion rate in units of 10$^{16} gs^{-1}$. These equations assume
that the disc is in the steady state. The best fits to the mass
accretion rate are shown in Table \ref{mdot}. These mass accretion
rates are too low to allow U Sco to gain mass given ejected mass
estimates of 10$^{-6}$-10$^{-7}$\,M$_{\odot}$ \citep{diaz,takei} and a recurrence period
of $\sim$10 years. 

An alternative measure of the mass accretion rate is the strength of
the He\,{\sc ii} 4686{\AA} line. \cite{patterson} show that the
strength of this line is empirically related to the mass accretion
rate in compact CVs; the mass accretion rates derived via this method
are shown in Table \ref{heacc}. These mass accretion rates are
$\sim$100 times higher than those derived from fitting the VLT
spectra and are high enough that U Sco could be gaining mass over each
outburst cycle. These mass accretion rates also show that the He\,{\sc
ii} 4686{\AA} line strength is consistent with a high mass white dwarf
accreting at a high rate without requiring an over-abundance of
helium, contrary to the suggestion of \cite{hachisu} that the
companion is helium-rich, and consistent
with the solar helium abundance found in the ejecta of the 2010
outburst \citep{maxwell}. 

The discrepancy between the two sets of mass accretion rates, and the
differences between the VLT and SALT spectra, is consistent with an increase in luminosity of the disc in the time between
the two sets of observations. \cite{schaeferdisc} find that the disc
has become fully re-established by 67 days after outburst since by
this time the system has returned to the pre-outburst magnitude level,
however here we find from the spectroscopic evolution until $\sim$30
months after outburst and the low mass accretion rates derived from
fits to the VLT spectra, which indicates that the disc is not in the
steady state at this time, that it takes at least
$\sim$18 months to fully reform.

\begin{table}
\centering
\begin{tabular}{ccc}\hline
Phase & $\dot{M} (M_{\odot}/yr)$ & T$_{companion}$\\ \hline
0.25 & 7.31$^{+0.48}_{-0.64} \times 10^{-9}$ & 5000$^{+750}_{-250}$ \\
0.40 & 6.52$^{+0.48}_{-0.48} \times 10^{-9}$& 6500$^{+250}_{-250}$ \\
0.43 & 7.31$^{+1.12}_{-0.48} \times 10^{-9}$& 7000$^{+250}_{-750}$ \\
0.46 & 7.31$^{+0.48}_{-0.48} \times 10^{-9}$& 6500$^{+250}_{-250}$ \\
0.62 & 4.45$^{+0.16}_{-0.32} \times 10^{-9}$& 6000$^{+250}_{-250}$ \\ \hline
\end{tabular}
\caption{Best fits for mass accretion rates and companion temperature. Errors are the 95\% confidence interval.}
\label{mdot}
\end{table}

\begin{table}
\centering
\begin{tabular}{ccc} \hline
Phase & L (ergs\,s$^{-1}$)& $\dot{M} (M_{\odot} yr^{-1})$ \\ \hline
0.25 &4.47$\pm0.12 \times10^{31}$& 2.81$\pm0.03 \times 10^{-7}$  \\
0.40 &11.1$\pm0.14 \times10^{31}$& 9.09$\pm 0.01 \times 10^{-7}$  \\
0.43 &9.34$\pm0.14 \times10^{31}$& 7.22$\pm 0.01 \times 10^{-7}$  \\
0.46 &9.43$\pm0.14 \times10^{31}$& 7.39$\pm 0.01 \times 10^{-7}$  \\
0.55 &2.56$\pm0.10 \times10^{31}$& 1.38$\pm 0.04 \times 10^{-7}$  \\
0.62 &8.42$\pm0.10 \times10^{31}$& 6.29$\pm 0.01 \times 10^{-7}$  \\
0.77 &4.12$\pm0.28 \times10^{31}$& 2.56$\pm 0.07 \times 10^{-7}$  \\ \hline
\end{tabular}
\caption{Mass accretion rates derived from He\,{\sc ii} 4686{\AA} flux.}
\label{heacc}
\end{table}

\section{The companion star}
The effective temperature of the companion star was constrained via
fitting the VLT spectra as described above. The effective temperatures
of the best fit models, and the derived luminosity, radius, and
spectral class, are shown in Table \ref{spectral}. The derived
luminosities and radii suggest that the companion is a sub-giant star,
in agreement with \cite{thoroughgood}, and that it fills its Roche
lobe. At orbital phases 0.40-0.62 the spectral class is determined to
be F2-G3 at the 95\% confidence level. The effect of irradiation from the hot component, causing
the inwards facing hemisphere of the companion to be heated, appears
to be present here. At phase 0.25 there is significant contribution from the outer, cooler
hemisphere of the companion resulting in a cooler effective temperature. 

\begin{table}
\centering
\begin{tabular}{cccccc}\hline
Phase & T$_{companion}$\,(K) &$L/L_{\odot}$&$R/R_{\odot}$ &Spectral class & Spectral class at 95\% \\ \hline
0.25 &5000$^{+750}_{-250}$ &1.41$^{+1.06}_{-0.24}$ &1.60$^{+0.77}_{-0.21}$& K0 & G3 - K2\\
0.40 &6500$^{+250}_{-250}$ &3.76$^{+0.67}_{-0.51}$ &1.54$^{+0.18}_{-0.16}$& F7 & F5 - F9\\
0.43 &7000$^{+250}_{-750}$ &4.32$^{+0.71}_{-1.56}$ &1.43$^{+0.40}_{-0.40}$& F3 & F2 - F9\\
0.46 &6500$^{+250}_{-250}$ &3.41$^{+0.60}_{-0.47}$ &1.47$^{+0.17}_{-0.15}$& F7 & F5 - F9\\
0.62 &6000$^{+250}_{-250}$ &4.19$^{+0.79}_{-0.62}$ &1.91$^{+0.24}_{-0.21}$& G0 & F9 - G3\\ \hline
\end{tabular}
\caption{Spectral type from fits to VLT data.}
\label{spectral}
\end{table}

\section{Conclusion}
From VLT and SALT spectra of U Sco taken $\sim$18 and $\sim$30 months
after outburst the accretion disc has been shown to take at least 18
months to become fully reformed. The spectral type of the companion
has been constrained to be F8$^{+5}_{-6}$ when the inwards facing
hemisphere of the companion is observed.

\bibliographystyle{asp2010}
\bibliography{maxwell}

\begin{thebibliography}{}
\expandafter\ifx\csname natexlab\endcsname\relax\def\natexlab#1{#1}\fi
\expandafter\ifx\csname url\endcsname\relax
  \def\url#1{\texttt{#1}}\fi
\expandafter\ifx\csname urlprefix\endcsname\relax\def\urlprefix{URL }\fi
\providecommand{\eprint}[2][]{\url{#2}}

\bibitem[{{Diaz} et~al.(2010){Diaz}, {Williams}, {Luna}, {Moraes}, \&
  {Takeda}}]{diaz}
{Diaz}, M.~P., {Williams}, R.~E., {Luna}, G.~J., {Moraes}, M., \& {Takeda}, L.
  2010, \aj, 140, 1860. \eprint{1009.4740}

\bibitem[{{Hachisu} et~al.(1999){Hachisu}, {Kato}, {Nomoto}, \&
  {Umeda}}]{hachisu}
{Hachisu}, I., {Kato}, M., {Nomoto}, K., \& {Umeda}, H. 1999, \apj, 519, 314.
  \eprint{arXiv:astro-ph/9902303}

\bibitem[{{Johnston} \& {Kulkarni}(1992)}]{johnston}
{Johnston}, H.~M., \& {Kulkarni}, S.~R. 1992, \apj, 396, 267

\bibitem[{{Kurucz}(1992)}]{kurucz}
{Kurucz}, R.~L. 1992, in The Stellar Populations of Galaxies, edited by
  B.~{Barbuy}, \& A.~{Renzini}, vol. 149 of IAU Symposium, 225

\bibitem[{{Maxwell} et~al.(2012){Maxwell}, {Rushton}, {Darnley}, {Worters},
  {Bode}, {Evans}, {Eyres}, {Kouwenhoven}, {Walter}, \& {Hassall}}]{maxwell}
{Maxwell}, M.~P., {Rushton}, M.~T., {Darnley}, M.~J., {Worters}, H.~L., {Bode},
  M.~F., {Evans}, A., {Eyres}, S.~P.~S., {Kouwenhoven}, M.~B.~N., {Walter},
  F.~M., \& {Hassall}, B.~J.~M. 2012, \mnras, 419, 1465. \eprint{1109.2468}

\bibitem[{{Patterson} \& {Raymond}(1985)}]{patterson}
{Patterson}, J., \& {Raymond}, J.~C. 1985, \apj, 292, 550

\bibitem[{{Schaefer} et~al.(2011){Schaefer}, {Pagnotta}, {LaCluyze},
  {Reichart}, {Ivarsen}, {Haislip}, {Nysewander}, {Moore}, {Oksanen},
  {Worters}, {Sefako}, {Mentz}, {Dvorak}, {Gomez}, {Harris}, {Henden}, {Guan
  Tan}, {Templeton}, {Allen}, {Monard}, {Rea}, {Roberts}, {Stein}, {Maehara},
  {Richards}, {Stockdale}, {Krajci}, {Sjoberg}, {McCormick}, {Revnivtsev},
  {Molkov}, {Suleimanov}, {Darnley}, {Bode}, {Handler}, {Lepine}, \&
  {Shara}}]{schaeferdisc}
{Schaefer}, B.~E., {Pagnotta}, A., {LaCluyze}, A.~P., {Reichart}, D.~E.,
  {Ivarsen}, K.~M., {Haislip}, J.~B., {Nysewander}, M.~C., {Moore}, J.~P.,
  {Oksanen}, A., {Worters}, H.~L., {Sefako}, R.~R., {Mentz}, J., {Dvorak}, S.,
  {Gomez}, T., {Harris}, B.~G., {Henden}, A.~A., {Guan Tan}, T., {Templeton},
  M., {Allen}, W.~H., {Monard}, B., {Rea}, R.~D., {Roberts}, G., {Stein}, W.,
  {Maehara}, H., {Richards}, T., {Stockdale}, C., {Krajci}, T., {Sjoberg}, G.,
  {McCormick}, J., {Revnivtsev}, M., {Molkov}, S., {Suleimanov}, V., {Darnley},
  M.~J., {Bode}, M.~F., {Handler}, G., {Lepine}, S., \& {Shara}, M.~M. 2011,
  \apj, 742, 113. \eprint{1108.1214}

\bibitem[{{Schaefer} \& {Ringwald}(1995)}]{schaefer95}
{Schaefer}, B.~E., \& {Ringwald}, F.~A. 1995, \apjl, 447, L45

\bibitem[{{Takei} et~al.(2013){Takei}, {Drake}, {Tsujimoto}, {Ness}, {Osborne},
  {Starrfield}, \& {Kitamoto}}]{takei}
{Takei}, D., {Drake}, J.~J., {Tsujimoto}, M., {Ness}, J.-U., {Osborne}, J.~P.,
  {Starrfield}, S., \& {Kitamoto}, S. 2013, ArXiv e-prints. \eprint{1303.5766}

\bibitem[{{Thoroughgood} et~al.(2001){Thoroughgood}, {Dhillon}, {Littlefair},
  {Marsh}, \& {Smith}}]{thoroughgood}
{Thoroughgood}, T.~D., {Dhillon}, V.~S., {Littlefair}, S.~P., {Marsh}, T.~R.,
  \& {Smith}, D.~A. 2001, \mnras, 327, 1323. \eprint{arXiv:astro-ph/0107477}

\bibitem[{{Warner}(2003)}]{warner}
{Warner}, B. 2003, {Cataclysmic Variable Stars}

\end{thebibliography}

\end{document}